\documentclass[preprint2]{aastex}
\shorttitle{NGC 1333 IRAS 4A Disks}
\shortauthors{Choi et al.}
\received{}

\usepackage{times}
\newcommand{\Jypb}{{Jy beam$^{-1}$}}
\newcommand{\kms}{\mbox{km s$^{-1}$}}
\newcommand{\Msun}{{$M_\odot$}}
\newcommand{\NHt}{{\rm NH$_3$}}
\newcommand{\Vlsr}{{$V_{\rm LSR}$}}
\usepackage{epsfig}
\newcommand{\ploteps}[2]{\centerline{\ifdim#2=0mm\epsfig{file=#1}%
                                     \else\epsfig{file=#1,width=#2}\fi}}

\slugcomment{To appear in the Astrophysical Journal}

\begin{document}
\fontsize{10}{10.6}\selectfont
\title{Ammonia Imaging of the Disks
       in the NGC 1333 IRAS 4A Protobinary System}
\author{\sc Minho Choi\altaffilmark{1},
            Ken'ichi Tatematsu\altaffilmark{2},
            Geumsook Park\altaffilmark{1,3},
            and Miju Kang\altaffilmark{1,3}}
\altaffiltext{1}{International Center for Astrophysics,
                 Korea Astronomy and Space Science Institute,
                 Hwaam 61-1, Yuseong, Daejeon 305-348, South Korea;
                 minho@kasi.re.kr.}
\altaffiltext{2}{National Astronomical Observatory of Japan,
                 2-21-1 Osawa, Mitaka, Tokyo 181-8588, Japan.}
\altaffiltext{3}{Department of Astronomy and Space Science,
                 Chungnam National University, Yuseong, Daejeon 305-764,
                 South Korea.}
\setcounter{footnote}{3}

\begin{abstract}
\fontsize{10}{10.6}\selectfont
The NGC 1333 IRAS 4A protobinary was observed
in the ammonia (2, 2) and (3, 3) lines and in the 1.3 cm continuum
with a high resolution (about 1.0$''$).
The ammonia maps show two compact sources, one for each protostar,
and they are probably protostellar accretion disks.
The disk associated with IRAS 4A2 is seen nearly edge-on
and shows an indication of rotation.
The A2 disk is brighter in the ammonia lines
but dimmer in the dust continuum than its sibling disk,
with the ammonia-to-dust flux ratios
different by about an order of magnitude.
This difference suggests
that the twin disks have surprisingly dissimilar characters,
one gas-rich and the other dusty.
The A2 disk may be unusually active or hot,
as indicated by its association with water vapor masers.
The existence of two very dissimilar disks in a binary system suggests
that the formation process of multiple systems
has a controlling agent lacking in the isolated star formation process
and that stars belonging to a multiple system
do not necessarily evolve in phase with each other.
\end{abstract}

\keywords{accretion disks --- binaries: general
          --- ISM: individual (NGC 1333 IRAS 4A) --- stars: formation}

\section{INTRODUCTION}

The NGC 1333 region in Perseus is a site of active star formation
producing a young cluster of Sun-like low-mass stars
(Aspin et al. 1994; Bally et al. 1996; Rodr{\'\i}guez et al. 1999).
IRAS 4A is one of the brightest submillimeter sources
among the deeply embedded protostars in NGC 1333 (Sandell \& Knee 2001).
The cold spectrum of dust emission suggests that IRAS 4A is very young
(Sandell et al. 1991),
and IRAS 4A drives molecular outflows with an interesting morphology
(Blake et al. 1995; Girart et al. 1999; Choi 2001).
Radio interferometric observations revealed
that there are two sources in IRAS 4A,
probably gravitationally bound to each other
(Lay et al. 1995; Looney et al. 2000).

In all the continuum images published so far,
the primary, A1, is more luminous than the secondary, A2,
by a factor of $\sim$3
(Looney et al. 2000; Reipurth et al. 2002; Girart et al. 2006),
which suggests that A1 may be more massive than A2.
However, recent observations in the SiO and the H$_2$ lines
puzzlingly showed
that the outflow driven by A2 (northeast-southwestern outflow)
is much more powerful and larger
than the one driven by A1 (southern outflow) (Choi 2005; Choi et al. 2006).
This inverse relation between the (dust) mass and the outflow activity
challenges conventional models of star formation
since both of them are supposed to be positively correlated
with the mass accretion rate (Cabrit \& Andr{\'e} 1991; Bachiller 1996).

To investigate the structure of the NGC 1333 IRAS 4A system,
we observed the region in several molecular tracers.
In this Letter, we present our high angular resolution observations
in the \NHt\ lines.
In \S~2 we describe our \NHt\ observations.
In \S~3 we report the results of the \NHt\ imaging
and discuss the physical process in the IRAS 4A system.

\section{OBSERVATIONS}

The NGC 1333 IRAS 4 region was observed using the Very Large Array (VLA)
of the National Radio Astronomy Observatory\footnote{
The NRAO is a facility of the National Science Foundation
operated under cooperative agreement by Associated Universities, Inc.}
in the \NHt\ (2, 2) and (3, 3) lines
(23722.6336 and 23870.1296 GHz, respectively)
and in the $\lambda$ = 1.3 cm continuum.
Twenty-five antennas were used
in the C-array configuration on 2004 March 5.
The continuum was observed for 20 minutes at the beginning
and for 10 minutes at the end of the observing track,
and the \NHt\ lines were observed
for 5 hours in the midsection of the track.
For each of the \NHt\ lines,
the spectral windows were set to have 64 channels
with a channel width of 0.049 MHz,
giving a velocity resolution of 0.62 \kms.
For the 1.3 cm continuum, 
the observations were made in the standard $K$-band continuum mode
(22.5 GHz or $\lambda$ = 1.33 cm).

The phase tracking center was ($\alpha$, $\delta$)
= (03$^{\mathrm h}$29$^{\mathrm m}$10.41$^{\mathrm s}$,
31\arcdeg13$'$32.2$''$) in J2000.0.
The nearby quasar 0336+323 (PKS 0333+321) was observed
to determine the phase and to obtain the bandpass response.
The flux was calibrated by observing the quasar 0713+438 (QSO B0710+439)
and by setting its flux density to 0.49 Jy,
which is the average of the flux density
measured within a day of our observations (VLA Calibrator Flux Density
Database\footnote{See http://aips2.nrao.edu/vla/calflux.html.}).
Comparison of the amplitude gave a flux density of 1.97 Jy for 0336+323,
and the flux uncertainty is $\sim$10\%.
To avoid the degradation of sensitivity owing to pointing errors,
pointing was referenced
by observing the calibrators at the $X$-band (3.6 cm).
This referenced pointing was performed
about once an hour and just before observing the flux calibrator.

Maps were made using a CLEAN algorithm.
The \NHt\ data produced synthesized beams
of FWHM = 0.97$''$ $\times$ 0.94$''$
and P.A. = 29\arcdeg\ for the (2, 2) line
and 0.98$''$ $\times$ 0.95$''$ and 47\arcdeg\ for the (3, 3) line,
when the imaging was done with a natural weighting.
The 1.3 cm continuum data produced a synthesized beam
of 1.02$''$ $\times$ 0.91$''$ and 71\arcdeg,
with a robust weighting.

\section{RESULTS AND DISCUSSION}

Figure 1 shows the \NHt\ (3, 3) line image
with an angular resolution similar to the SiO outflow image of Choi (2005).
While the \NHt\ map shows clumpy structures around the outflows,
the most striking feature is the compact structure
associated with the central objects,
which were not seen in the SiO map.
(The outflows will be discussed separately in a future paper.)
Figure~2 compares the \NHt\ maps with the 1.3 cm continuum map,
Table 1 lists the continuum parameters,
and Figure 3 shows the \NHt\ spectra.
The deconvolved FWHM sizes of the \NHt\ (3, 3) sources are
350 $\times$ 260 AU$^2$ (1.1$''$ $\times$ 0.8$''$) for A1
and 260 $\times$ 60 AU$^2$ (0.8$''$ $\times$ 0.2$''$) for A2,
assuming a distance of 320 pc (de Zeeuw et al. 1999).
Their compact nature suggests that the \NHt\ lines
are tracing accretion disks.
This interpretation is especially strong in the case of A2
because the emission structure is elongated
in the direction perpendicular to the main bipolar outflow.
The position angle difference
between the minor axis of the A2 disk and the outflow axis
is $\sim$20\arcdeg.
In addition, the blue/redshifted emission peaks of A2 are
displaced in a way suggestive of a rotating disk (Fig. 2$c$).

\begin{figure}[t]
\ploteps{nh3sio-lowres.eps}{88mm}
\vspace{-0.5\baselineskip}
\centerline{\scriptsize [See http://minho.kasi.re.kr/Publications.html
for the original high-quality figure.]}
\vspace{-0.5\baselineskip}
\caption{\small\baselineskip=0.825\baselineskip
Maps of the \NHt\ (3, 3) line ({\it contours})
and the SiO $v$ = 0 $J$ = 1 $\rightarrow$ 0 line ({\it gray scale})
toward the NGC 1333 IRAS 4A region.
The \NHt\ line intensity was
averaged over the velocity intervals of \Vlsr\ = (--1.0, 14.4) \kms,
which includes the line core and wings.
The \NHt\ map was convolved to have an angular resolution of FWHM = 2.0$''$,
and the rms noise is 0.20 m\Jypb.
The contour levels are 0.7, 1.0, 1.3, and 1.6 m\Jypb.
Dashed contours are for negative levels.
The SiO map is the same as the one shown in Fig. 3$a$ of Choi (2005).
{\it Plus signs}:
3.6 cm continuum sources (Reipurth et al. 2002).}
\end{figure}

\begin{figure*}[!t]
\ploteps{nh3kbr-lowres.eps}{185mm}
\vspace{-0.5\baselineskip}
\centerline{\scriptsize [See http://minho.kasi.re.kr/Publications.html
for the original high-quality figure.]}
\vspace{-0.5\baselineskip}
\caption{\small\baselineskip=0.825\baselineskip
Maps of the \NHt\ lines ({\it contours})
and the 1.3 cm continuum ({\it color scale}).
The color scale starts from 0.036 m\Jypb.
Shown at the bottom right corner are the synthesized beams.
{\it Plus signs}:
3.6 cm continuum sources (Reipurth et al. 2002).
{\it Filled circles}:
H$_2$O maser spots (Park \& Choi 2007).
($a$)
Map of the \NHt\ (2, 2) line core.
The line intensity was averaged
over the velocity interval of \Vlsr\ = (4.5, 8.9) \kms.
The contour levels are 1, 2, 3, and 4 times 0.8 m\Jypb,
and the rms noise is 0.27 m\Jypb.
Dashed contours are for negative levels.
($b$)
Map of the \NHt\ (3, 3) line core.
Map parameters are the same as ($a$).
($c$)
Maps of the blueshifted and the redshifted emission
of the \NHt\ (3, 3) line,
averaged over the velocity intervals
of \Vlsr\ = (3.9, 5.8) and (7.6, 9.5) \kms, respectively. 
The contour levels are 2, 3, 4, and 5 times 0.6 m\Jypb.
The straight line near the bottom right corner
corresponds to 500 AU at a distance of 320 pc.}
\end{figure*}

\begin{deluxetable}{p{13mm}cccccc}
\tabletypesize{\small}
\tablecaption{NGC 1333 IRAS 4 Continuum Source Parameters}%
\tablewidth{0pt}
\tablehead{
& \multicolumn{2}{c}{\sc Peak Position}
&& \multicolumn{2}{c}{\sc Flux Density\tablenotemark{a}} \\
\cline{2-3} \cline{5-6}
\colhead{\sc Source}
& \colhead{$\alpha_{\rm J2000.0}$} & \colhead{$\delta_{\rm J2000.0}$}
&& \colhead{Peak} & \colhead{Total} & \colhead{$\alpha$\tablenotemark{b}} }%
\startdata
A1\dotfill & 03 29 10.53 & 31 13 31.0 &
           & 1.44 $\pm$ 0.01 & 1.79 $\pm$ 0.03 & 3.3 \\
A2\dotfill & 03 29 10.41 & 31 13 32.6 &
           & 0.19 $\pm$ 0.01 & 0.25 $\pm$ 0.02 & 3.7 \\
BI\dotfill & 03 29 12.00 & 31 13 08.3 &
           & 0.36 $\pm$ 0.01 & 0.36 $\pm$ 0.01 & 3.7 \\
\enddata\\
\tablecomments{Units of right ascension are hours, minutes, and seconds,
               and units of declination are degrees, arcminutes,
               and arcseconds.}%
\tablenotetext{a}{Flux densities at 1.3 cm in mJy,
                  corrected for the primary beam response.}%
\tablenotetext{b}{Spectral index between 1.3 cm and 2.7 mm
                  (Looney et al. 2000).
                  The uncertainty in $\alpha$ is 0.1,
                  assuming that the uncertainty in the absolute flux scale
                  is 10 \%.}%
\end{deluxetable}

\subsection{Flux Anti-correlation}

Comparisons of the \NHt\ and the continuum maps
reveal a remarkable anti-correlation:
A1 is brighter than A2 in the continuum map,
but it is in the other way around in the \NHt\ maps.
For a quantitative analysis,
we may define an \NHt-to-dust flux ratio
$R = F({\rm NH}_3) / F({\rm 2.7\ mm})$,
which is the ratio of the total flux densities
of the \NHt\ (3, 3) line to the 2.7 mm continuum.
The flux densities from the \NHt\ data shown in Figure 2$b$
and the 2.7 mm data of Looney et al. (2000)
give $R$(A1) = (7.1 $\pm$ 1.1) $\times$ 10$^{-3}$
and $R$(A2) = (4.8 $\pm$ 0.7) $\times$ 10$^{-2}$.
(The uncertainties were estimated
by assuming a 10\%\ uncertainty in flux scales.)
That is, the \NHt-to-dust flux ratio of A2 is larger
than that of A1 by a factor of 6.8 $\pm$ 1.4.
(If we use the 1.3 cm flux densities,
$R$(A2)/$R$(A1) is about 12.)
Before making interpretations of the flux ratio as column density ratio,
several factors affecting the flux densities should be considered.

First, can the line optical depth affect the flux ratio?
Since the satellite hyperfine components
were not covered in our observations,
the optical depth cannot be deduced directly.
The line profiles (Fig. 3), however, provide useful information.
The spectra toward A2 have larger line widths and higher intensities
than those of A1,
which suggests that A2 may have relatively larger optical depths.
In addition, the (2, 2) spectrum toward A2 shows a self-absorption dip,
suggesting that the line may be optically thick.
Therefore, if the line optical depth effects are considered,
the degree of the anti-correlation would be severer.

Note that, however,
there are alternative explanations for the line profiles.
The line width difference can be explained
if source A2 is a nearly edge-on disk
and A1 is either a relatively more face-on disk or a static core.
Also, the central dip of the spectra can be caused
by a missing flux problem owing to large-scale structures.
This issue can be addressed by future observations
either with a higher angular resolution
or with a spectral coverage wide enough
to include the satellite hyperfine components.

Second, can the high \NHt\ flux of A2 be caused by a peculiar excitation?
The ratio between the (3, 3) and the (2, 2) lines
is not very useful for a quantitative analysis
because they belong to different species,
ortho-\NHt\ and para-\NHt, respectively (Ho \& Townes 1983).
Even so, the (3, 3) to (2, 2) line ratio
is similar in both sources (Fig. 3),
suggesting that they have similar excitation conditions.
In fact, the critical density of the \NHt\ inversion transitions
are low ($\sim$2 $\times$ 10$^3$ cm$^{-3}$; Ho \& Townes 1983),
and the \NHt\ molecules are expected to be thermalized.
Therefore, it is unlikely
that the difference in the \NHt-to-dust flux ratio
is caused by a peculiar excitation condition of \NHt\
in one of the sources.

Finally, can the dust properties affect the flux ratio?
The spectral index in principle
provides some information on the dust opacity index.
The spectral index of A1 is slightly lower than that of A2 (Table 1),
but these values cannot be used directly
because the contributions from free-free emission
to the 1.3 cm fluxes are not known.
Flux measurements at submillimeter wavelengths
are desirable for estimating the dust opacity index.
Girart et al. (2006) presented a 345 GHz map
that resolves the continuum peaks,
but their beam size was not quite small enough
for measuring the total flux densities of each source separately.
Therefore, this issue cannot be resolved with the currently available data,
and we presume that the dust opacity index of the two sources are similar.
Another variable that can affect the flux ratio is the dust optical depth.
The dust is most likely optically thin at centimeter wavelengths,
but it could be optically thick at submillimeter.
While the measurement of the optical depth is not easy,
if this effect is significant,
it would affect the stronger source, A1, more severely.
Therefore, a correction for the dust optical depth, if necessary,
would also make the degree of the anti-correlation severer.

\begin{figure}[t]
\ploteps{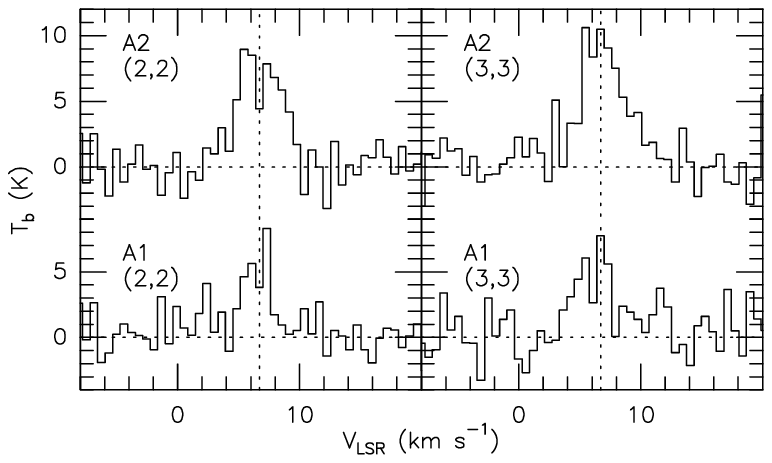}{88mm}
\caption{\small\baselineskip=0.825\baselineskip
Spectra of the \NHt\ lines
at the peak positions of the 3.6 cm continuum emission (see Fig. 2).
Contribution from the dust emission was subtracted
by determining the continuum flux level
in the velocity intervals of
(--6.9, --0.7) and (13.5,19.7) \kms\ for the (2, 2) line,
and (--9.9, 0.6) and (18.4, 23.3) \kms\ for the (3, 3) line.
{\it Vertical dotted lines}:
Systemic velocity of the IRAS 4A core
(\Vlsr\ = 6.7 \kms; Blake et al. 1995; Choi 2001).}
\end{figure}

In summary, the results of our observations suggest
that the IRAS 4A system contains
two sources of contrasting conditions.
The \NHt-to-dust flux ratio of A2 is $\sim$7 times larger than that of A1,
and the \NHt-to-dust column density ratios
may be different by a similar factor,
or by a larger factor
if the optical depth of the line or the dust emission is considered.
The difference between the two sources is huge,
considering that they are accreting matter
from a common protostellar envelope,
and must be caused by the physical and chemical processes
happening in each component.

\subsection{Possible Explanations}

There are two possible explanations
for the flux/column-density anti-correlation of the IRAS 4A system.
The main difference between them is the evolutionary status of A1.
While A2 is almost certainly a protostar,
the nature of A1 is less certain.

(1)
A1 and A2 are roughly coeval,
the \NHt\ lines trace two accretion disks,
and the anti-correlation is caused
by a peculiar physical/chemical condition in one of them.
In this scenario,
A2 is a protostar driving the northeast-southwestern outflow,
and A1 is another protostar driving the southern outflow.
(We will elaborate on this model in \S~3.3.)

(2)
A1 and A2 are in quite different stages of evolution,
the \NHt\ maps show
a (spherical) dense core (A1) and an accretion disk (A2),
and the anti-correlation is
an indication of the difference in their nature.
In this scenario, A2 is an actively accreting protostar,
and A1 is a pre-protostellar object without an outflow activity.
The strong millimeter continuum of A1 suggests a high concentration of dust.
Nevertheless, the compact structure detected by interferometers
suggests that A1 may not be a usual pre-protostellar object, either.
Then A1 could be a transitionary object,
either a pre-protostellar object on the verge of collapse
or a protostar immediately after the onset of collapse.

We prefer the first explanation for several reasons.
First, A1 is bright in the centimeter continuum (Reipurth et al. 2002).
This free-free emission is a clear sign of outflow activity.
Second, comparison of the mass estimates over a range of size scale
indicates that A1 has a steep density gradient.
[For example, the mass of the IRAS 4A envelope
within a diameter of $\sim$9$''$ is 3.2~\Msun,
when scaled to the distance of 320 pc (Sandell \& Knee 2001).
The mass of A1 within a $\sim$2.5$''$ box is 1.9 \Msun,
when scaled to 320 pc and scaled by the flux ratio between A1 and A2
(Looney et al. 2000).
Then the density of A1 (average within a diameter of 820 AU) is higher
than the density of the envelope (average within 2900 AU)
by a factor of 26 $\pm$ 4.]
Third, in the second scenario,
the southern outflow might be driven by an unknown protostellar object,
which is unlikely because NGC 1333 IRAS 4 is
one of the most extensively observed regions
of star formation near the Sun.
Therefore, we suppose that both A1 and A2 are protostars.
In the following section,
we will discuss the implications of our observations
based on the first explanation.

\subsection{Peculiarity of the IRAS 4A2 Disk}

The difference in the column density ratio suggests
that one of the IRAS 4A disks is peculiar.
There are several lines of evidence
indicating that the A2 disk is unusually active.
First, most of the water maser spots in this region
are intimately associated with A2 (Fig. 2),
and their velocities are close to the systemic velocity of the cloud core
(Furuya et al. 2003; Park \& Choi 2007),
suggesting the existence of shocked gas in the A2 disk.
Second, the outflow driven by A2 (northeast-southwestern bipolar outflow)
is stronger than the one driven by A1 (southern outflow)
(Choi 2005; Choi et al. 2006),
also suggesting that the outflow engine (the accretion disk)
is more active in A2 than in A1.
The northeast-southwestern outflow of IRAS A2
is one of the best collimated molecular outflows (Blake et al. 1995).
Third, A2 seems to have an unusually large $R$ value.
Comparison with other protostars may tell
which disk is the peculiar one.
Examples of protostellar disks detected in the \NHt\ (3, 3) line is rare,
but fortunately IRAS 4BI was detected.
IRAS 4BI is a single protostar located within the field of view
of our observations (Fig. 1).
Measurements of the flux densities
give $R$(BI) = (4 $\pm$ 2) $\times$ 10$^{-3}$,
which is similar to $R$(A1)
and suggests that A2 may be the abnormal one.
Therefore, the A2 disk may be unusually gas-rich or dust-poor.
Such a condition may be possible if the disk is very active or hot
so that the dust grains infalling from the protostellar envelope
may be destroyed and converted to gaseous molecules,
probably via evaporation of molecules in the grain mantle.

The anti-correlation between the gas and the dust flux densities
of the IRAS 4A disks
has important implications on the star formation process.
The standard models of accretion
are based on the cases of single central star,
and the mass accretion rate is
mainly related with the density structure of the protostellar envelope
(Shu et al. 1987).
In binary systems, however, as the IRAS 4A system shows,
even though the two components share a common envelope,
each of them can evolve in a distinctive way.
That is, there is an important controlling agent
that may be lacking in the case of isolated star formation.
A possibly crucial factor can be
the distribution and (mis-)alignment of angular momentum
(Bodenheimer 1995).
If the mass outflow rate is a good indicator of the mass accretion rate,
A2 may be growing much faster than A1,
by accreting matter through an active disk.

One interesting issue to be addressed in the future is the \NHt\ abundance.
Does the high column-density ratio of A2 mean
an overall enhancement of gaseous molecules relative to dust?
Or does it mean a selective enhancement
of \NHt\ (and related species) only?
Estimating the degree of enhancement is difficult
because comparison with CO lines cannot be interpreted easily
owing to the complicated chemistry of nitrogen-bearing molecules
(Charnley 1997)
and confusion with outflows.
Currently there is no reliable estimates of \NHt\ abundance
in protostellar disks,
and it is needed to make high-resolution images of the two disks
in a variety of molecular lines.

Since planets do exist in multiple-star systems (Raghavan et al. 2006),
our results have interesting implications on the planet formation.
If the A2 disk is indeed gas-rich/dust-poor,
and if such a condition can persist
until the planet-forming phase of the disk evolution,
planetary systems produced in such disks
may look very different from our solar system.
We may speculate that, for example, such a system may strongly favor
the growth of gas-giant planets.
Thus, multiple-protostar systems can produce
diverse types of planetary systems.

\acknowledgements

We thank J. Cho, K.-T. Kim, and Y. Lee
for helpful discussions and encouragement.
G. P. was partially supported by the Brain Korea 21 project
of the Korean Government.


\end{document}